\documentstyle[prd,preprint,tighten,aps,eqsecnum,%
amssymb,newlfont,epsfig]{revtex}
\newcommand{\kkk}{$K^0 \bar K^0$}
\begin{document}
\preprint{
\begin{tabular}{r}
UWThPh-1999-9\\
February 1999
\end{tabular}
}
\title{Quantum mechanics, Furry's hypothesis and a measure of decoherence 
in the \kkk\ system}
\author{R.A. Bertlmann, W. Grimus and B.C. Hiesmayr}
\address{Institut f\"ur Theoretische Physik,
Universit\"at Wien\\
Boltzmanngasse 5,
A-1090 Vienna, Austria}
\maketitle
\begin{abstract}
We consider strangeness correlations of the EPR type in 
\kkk\ pairs created in a
$J^{PC} = 1^{--}$ state as a function of time 
under the hypothesis that spontaneous 
decoherence takes place. We parameterize the degree of
decoherence by a factor $(1-\zeta)$ which multiplies the
quantum-mechanical interference terms occurring in the
amplitudes for like and unlike strangeness events and discuss the
dependence of this procedure on the basis chosen in the
$K^0$--$\bar{K}^0$ space to which the interference terms
correspond. Consequently, all statements about the
``decoherence parameter'' $\zeta$ inferred from experimental
data are basis-dependent as well. We illustrate this point 
by estimating the value of $\zeta$ for the two bases $\{K_L, K_S\}$ and
$\{K^0, \bar{K}^0\}$ with the help of recent data of the CPLEAR experiment.
\end{abstract}

\section{Introduction}
Since the first formulation of the EPR paradox in 1935 
\cite{EPR} tests of quantum mechanics (QM) against local realistic 
(hidden variable) theories are of great interest. 
According to QM, if a pair of particles is created by any kind of
interaction in an entangled state, the two particle wave function 
retains its non-separable character even
if the particles are space-like separated. The famous inequality
of J.S.\ Bell \cite{bell} made it possible 
to discriminate quantitatively between the predictions of QM and of local 
realistic theories.
Numerous experiments have been performed since, in particular,
using entangled photon states, which all confirmed QM (see,
e.g., Refs. \cite{aspect,kwiat}). On the particle physics side,
entangled \kkk\ and $B^0 \bar{B}^0$ states are suitables objects for
the study of EPR-like correlations \cite{lee,six,selleri,datta}. 
The presence of the QM interference term can also be deduced 
from existing data on $B^0 \bar{B}^0$ systems produced in 
the decay of $\Upsilon(4S)$ as was demonstrated in Refs.
\cite{BG97,dass,BG98}. Recently, an important experiment of the
CPLEAR Collaboration using strangeness correlations of \kkk\
pairs created by annihilation of $p \bar p$ pairs at rest
showed impressively the non-separability of the QM \kkk\ wave
function in this situation \cite{CPLEAR98}. This experiment
excluded a spontaneous wave function factorization immediately
after the \kkk\ creation  with a CL of more than 99.99\%.

In this paper we want to draw the attention to the fact that the
notion of spontaneous factorization depends on the basis chosen
in the two systems which, after the creation of the particle
pairs, are non-interacting. The reason for this is that 
a basis has to be fixed with
respect to which the process of spontaneous factorization takes
place \cite{furry} (see also Refs. \cite{dass,BG98}). 
In our case of interest the two
non-interacting systems are given by the two neutral kaons
moving into opposite directions in the rest frame of the \kkk\ source.
In this paper we discuss spontaneous factorization and decoherence
of the \kkk\ wave function created by a $J^{PC} = 1^{--}$ state
as a function of the basis chosen to describe the
non-interacting neutral kaons on each side of the source. 
We modify the quantum-mechanical interference term of the
entangled 2-kaon state by multiplying
it with $(1-\zeta)$.
Thus we change QM expressions in such a way
that the decoherence parameter $\zeta$ \cite{eberhard} parameterizes
the deviation from QM (corresponding to $\zeta = 0$) 
and gives a measure how far the total system is
away from total decoherence or spontaneous factorization ($\zeta
= 1$). As an observable relevant in this discussion we use the time-dependent
asymmetry $A(t_r,t_l)$ of like and unlike strangeness events measured
by the CPLEAR Collaboration \cite{CPLEAR98} and discussed in
their paper. The proper times at which the strangeness of the
neutral kaons is measured at the right and the left of the \kkk\
source at rest are denoted by $t_r$ and $t_l$, respectively. We
investigate the dependence of $A(t_r,t_l)$ on the basis chosen in
the 2-dimensional $K^0$--$\bar{K}^0$ space 
and on the decoherence parameter $\zeta$.
Then we use the results on $A(t_r,t_l)$ 
of the CPLEAR experiment \cite{CPLEAR98}
to estimate the value of the decoherence parameter for two basis
choices: first we consider the basis given by $\{K_L, K_S\}$,
which was used in Ref. \cite{CPLEAR98} to compare spontaneous
factorization of the wave function with the result of QM, and,
secondly, the basis $\{K^0, \bar{K}^0\}$. As we will see,
numerically the two estimates of $\zeta$ will be quite different
and, therefore, the statement that spontaneous factorization is
excluded with a CL of more than 99.99\% is valid only for the
$\{K_L$, $K_S\}$ basis in the factorization process. Also we will
see that this basis is the only one which leads to a vanishing
asymmetry upon spontaneous factorization.

Let us mention that strangeness in the \kkk\ system is a
quantity analogous to 
polarization in the well-known entangled two-photon systems. 
However, strangeness is time-dependent due to
$K^0 \leftrightarrow \bar K^0$ oscillations and it has been shown
that because of the actual values of the decay constants of 
$K_L$ and $K_S$ it is not possible 
to find an experimental set-up for violating an inequality of
the Bell type \cite{ghirardi91,ghirardi92,trixi} by using the
strangeness of the kaons. This is different from situations where Bell
inequalities for kaon systems can be tested with kaon decay products
\cite{bellineq}. Finally, we want to stress that the
\kkk\ system considered here is a further example exhibiting QM
interference effects over macroscopic distances with the
neutral kaons being separated several centimeters when their
strangeness is measured in the CPLEAR experiment.

\section{A formalism for the \kkk\ system in an arbitrary basis and
with modified interference terms}\label{theoretical basis choice}

A \kkk\ pair, created in a $J^{PC}=1^{--}$ quantum state and thus 
antisymmetric under C and P, is described at proper times 
$t_r=t_l=0$ by the an entangled state:
\begin{equation}\label{1}
| \psi (0,0) \rangle =\frac{1}{\sqrt{2}}
\left\{ | K^0 \rangle_r \otimes\! | \bar K^0 \rangle _l - 
| \bar K^0 \rangle _r \otimes | K^0 \rangle _l \right\} \,.
\end{equation}
Here $r$ and $l$ denote the neutral kaons on the right and left side
of the source. On the other hand,
the physical states which decay are the long and short-lived states:
\begin{eqnarray}\label{2}
| K_S \rangle & = & p | K^0 \rangle - q | \bar K^0 \rangle \,,
\nonumber\\
| K_L \rangle & = & p | K^0 \rangle + q | \bar K^0 \rangle \,,
\end{eqnarray}
where the normalization $|p|^2 + |q|^2 = 1$ is understood.
Rewriting the initial state (\ref{1}) in the basis $\{K_L, K_S\}$, we
obtain 
\begin{equation}\label{3}
| \psi (0,0) \rangle  = \frac{1}{2 \sqrt{2}\, p q}
\left\{ | K_S \rangle_r \otimes | K_L \rangle_l - 
        | K_L \rangle_r \otimes | K_S \rangle_l \right\} \,.
\end{equation}
Of course, with respect to QM the states (\ref{1}) and (\ref{3}) are 
identical and will lead to equal probabilities. But if we modify 
interference terms by introducing the decoherence
parameter $\zeta$ the derived probabilities
depend on the basis chosen \cite{furry}, a feature discussed 
already for the analogously entangled
$B^0 \bar B^0$ state in Refs. \cite{BG97,dass,BG98}.

To develop a general formalism for the neutral kaons we 
take an arbitrary basis \cite{BG98}
\begin{eqnarray}\label{3.5}
| k_j \rangle  = S_{1j} | K^0 \rangle  + S_{2j} | \bar
K^0 \rangle  \quad \mathrm{with} \quad j=1,2,
\end{eqnarray}
where $S_{ij}$ are elements of an arbitrary invertible matrix $S$. 
Thus the two special
cases considered above correspond to $S=\mathbf{1}$ and $S=M$ with
\begin{equation}
M =
\left( \begin{array}{lr}
p & p\\
q & -q\\
\end{array} \right) \,,
\end{equation}
respectively.

According to the Wigner--Weisskopf approximation the decaying states 
evolve exponentially in time:
\begin{eqnarray}
| K_S (t)\rangle  &=& g_S(t) | K_S \rangle \,, \nonumber\\
| K_L (t)\rangle  &=& g_L(t) | K_L \rangle 
\end{eqnarray}
with 
\begin{equation}
g_{S,L}(t) = e^{-i \lambda_{S,L} t} \quad \mathrm{and} \quad
\lambda_{S,L} = m_{S,L} - \frac{i}{2} \Gamma_{S,L} \,.
\end{equation}
The subsequent time evolution for $K^0$ and $\bar K^0$ is
therefore given by
\begin{eqnarray}
| K^0(t) \rangle  &=& 
g_{+}(t) | K^0 \rangle  + \frac{q}{p} g_{-}(t) | \bar K^0 \rangle \,,
\nonumber\\
| \bar K^0(t) \rangle  &=& 
\frac{p}{q} g_{-}(t) | K^0 \rangle + g_{+}(t) | \bar K^0 \rangle 
\end{eqnarray}
with
\begin{equation}
g_{\pm}(t)= \frac{1}{2} \left[ \pm e^{-i \lambda_S t} + e^{-i \lambda_L t} 
\right] \,.
\end{equation}
Defining the time evolution matrix by
\begin{equation}
T(t) \equiv M \hat g(t) M^{-1} S \quad \mathrm{with} \quad 
\hat g = \mathrm{diag} (g_L,g_S)
\end{equation}
and
\begin{equation}
M \hat g(t) M^{-1} =
\left( \begin{array}{cc}
g_{+}(t) & \frac{p}{q} g_-(t)\\
\frac{q}{p} g_-(t) & g_+(t)
\end{array} \right)
\end{equation}
we can write the time evolution of the basis vectors (\ref{3.5})
as
\begin{equation}
| k_j(t) \rangle  = 
T_{1j}(t) | K^0 \rangle  + T_{2j}(t) | \bar K^0 \rangle  
\quad \mathrm{for} \quad j=1,2.
\end{equation}
Then the initial state given by (\ref{1}) or (\ref{3}), written
in terms of an arbitrary basis (\ref{3.5}) as
\begin{equation}
| \psi(0,0) \rangle = \frac{1}{\sqrt{2} \det S} 
\left\{ | k_1 \rangle_r \otimes | k_2 \rangle_l - 
        | k_2 \rangle_r \otimes | k_1 \rangle_l \right\} \,,
\end{equation}
has the time evolution
\begin{equation}\label{timeevo}
| \psi(t_r,t_l) \rangle = \frac{1}{\sqrt{2} \det{S}} 
\left\{ T_{i1}(t_r) T_{j2}(t_l) - T_{i2}(t_r) T_{j1}(t_l) \right\} 
| \hat k_i \rangle \otimes | \hat k_j \rangle \,,
\end{equation}
where we have defined
\begin{equation}
| \hat k_1 \rangle \equiv | K^0 \rangle , \quad 
| \hat k_2 \rangle \equiv | \bar K^0 \rangle \,.
\end{equation}
In Eq.(\ref{timeevo}) a summation over equal indices is understood.
At this point we want to stress that QM is invariant under
the basis manipulations we have made up to now, as has been explicitely
demonstrated in the case of the $B^0 \bar B^0$ system in Ref. \cite{BG98}.

Now we are ready to modify QM. The class of observables we are
interested in is the probability that the state $\psi$ evolves 
into final states $f_1$ and $f_2$, which are measured at proper times 
$t_r$ on the right side and $t_l$ on the left side of the \kkk\ source. 
This probability is given by
\begin{eqnarray}
& &
\left| \langle f_1 \otimes f_2 |
\psi(t_r,t_l) \rangle \right|^2 = \frac{1}{2 | \det{S} |^2} 
\nonumber\\
& & \hphantom{AA}
\times \left\{ 
\left| \langle f_1 | k_1(t_r) \rangle \right|^2 
\left| \langle f_2 | k_2(t_l) \rangle \right|^2 +
\left| \langle f_1 | k_2(t_r) \rangle \right|^2 
\left| \langle f_2 | k_1(t_l) \rangle \right|^2 \right. \nonumber\\
& & \hphantom{AA}
-2\: (1-\zeta) \left. \mathrm{Re} \left( 
       \langle f_1 | k_1(t_r) \rangle ^*
       \langle f_2 | k_2(t_l) \rangle ^*
       \langle f_1 | k_2(t_r) \rangle 
       \langle f_2 | k_1(t_l) \rangle \right) \right\} \,,
\label{ff}
\end{eqnarray}
where the usual QM interference term has been modified by the 
factor $1-\zeta$. QM corresponds to $\zeta =0$ and for $\zeta = 1$ 
no interference term is present, corresponding to Furry's hypothesis
of spontaneous factorization \cite{furry} (called ``method A''
in his paper). The aim of our investigation is to find the range
of $\zeta$ allowed by present experimental data.
We will take the results of experiment of the CPLEAR Collaboration 
\cite{CPLEAR98} where
they measured four final states of the neutral K-meson pairs 
$(f_1,f_2)=(K^0,K^0)$, $(\bar K^0,\bar K^0)$, $(K^0,\bar K^0)$, 
$(\bar K^0, K^0)$. With these states we obtain the
following probabilities.\\
\textbf{Like-strangeness events:} 
final states $(K^0, K^0)$ and $(\bar K^0, \bar K^0)$
\begin{eqnarray}
P^S_\zeta(K^0, t_r; K^0, t_l) &=& 
\frac{1}{8} \left| \frac{p}{q} \right|^2 P_\mathrm{like}^\mathrm{QM}(t_r,t_l)
+ \zeta \frac{1}{| \det{S}|^2} 
\mathrm{Re} \left\{ 
(T_{11}^*(t_r) T_{12}(t_r)) (T_{11}^*(t_l) T_{12}(t_l))^*
\right\} \,, \nonumber\\
P^S_\zeta(\bar K^0, t_r; \bar K^0, t_l) &=& 
\frac{1}{8} \left| \frac{q}{p} \right|^2 P_\mathrm{like}^\mathrm{QM}(t_r,t_l)
+ \zeta \frac{1}{| \det{S}|^2} 
\mathrm{Re} \left\{ 
(T_{22}^*(t_r) T_{21}(t_r))^* (T_{22}^*(t_l) T_{21}(t_l))
\right\} \,. \nonumber\\
&&\label{3.3}
\end{eqnarray}
\textbf{Unlike-strangeness events:} 
final states $(K^0, \bar K^0)$ and $(\bar K^0, K^0)$
\begin{eqnarray}
P^S_\zeta(K^0, t_r;\bar K^0, t_l) &=&  
\frac{1}{8} P_\mathrm{unlike}^\mathrm{QM}(t_r,t_l)
+ \zeta \frac{1}{| \det{S}|^2} 
\mathrm{Re} \left\{ 
T_{11}^*(t_r) T_{12}(t_r) T_{22}^*(t_l) T_{21}(t_l)
\right\} \,, \nonumber\\
P^S_{\zeta}(\bar K^0, t_r; K^0, t_l) &=& 
\frac{1}{8} P_\mathrm{unlike}^\mathrm{QM}(t_r,t_l)
+ \zeta \frac{1}{| \det{S}|^2} 
\mathrm{Re} \left\{ 
T_{22}^*(t_r) T_{21}(t_r) T_{11}^*(t_l) T_{12}(t_l)
\right\} \,.
\label{3.4}
\end{eqnarray}
The QM probabilities in Eqs.(\ref{3.3}) and (\ref{3.4}), apart
from CP-violating effects, are given by \cite{six}
\begin{eqnarray}
P_\mathrm{like}^\mathrm{QM} &=& 
2\, e^{-\Gamma(t_r+t_l)}
\left\{ \cosh \left( {\textstyle \frac{1}{2}} 
\Delta \Gamma \Delta t \right) -
\cos(\Delta m\, \Delta t) \right\} \,, \nonumber\\
P_\mathrm{unlike}^\mathrm{QM} &=& 
2\, e^{-\Gamma(t_r+t_l)}
\left\{ \cosh \left( {\textstyle \frac{1}{2}} 
\Delta \Gamma \Delta t \right) +
\cos(\Delta m\, \Delta t) \right\} 
\label{QMterms}
\end{eqnarray}
with 
\begin{equation}
\Delta t \equiv t_r - t_l, \quad \Delta m \equiv m_L-m_S, \quad
\Delta \Gamma = \Gamma_L - \Gamma_S
\quad \mathrm{and} \quad \Gamma = \frac{1}{2}(\Gamma_L +\Gamma_S) \,.
\end{equation}
In contrast to the QM terms (\ref{QMterms}), the ``decoherence terms''
proportional to $\zeta$ in Eqs.(\ref{3.3}) and (\ref{3.4}) depend
on the matrix $S$, ie., on the choice of the basis.
These terms can be characterized by the two functions
\begin{eqnarray}
T_{11}^*(t) T_{12}(t) &=& 
I_+(t) S_{11}^* S_{12} + I_-(t) \left| \frac{p}{q} \right|^2
S_{21}^* S_{22} +I_{+-}(t) \frac{p}{q} S_{11}^* S_{22} +
I_{-+}(t) \frac{p^*}{q^*} S_{21}^* S_{12} \,, \nonumber\\
T_{22}^*(t) T_{21}(t) &=& 
I_+(t) S_{22}^* S_{21} + I_-(t) \left| \frac{q}{p} \right|^2
S_{12}^* S_{11} +I_{+-}(t) \frac{q}{p} S_{22}^* S_{11} +
I_{-+}(t) \frac{q^*}{p^*} S_{12}^* S_{21} \,,
\label{T}
\end{eqnarray}
where the t-dependent functions are defined by
\begin{eqnarray}
I_{\pm}(t) &=& | g_{\pm}(t) |^2 =
\frac{1}{2}\, e^{-\Gamma t}
\left\{ \cosh \left( {\textstyle \frac{1}{2}} 
\Delta \Gamma\, t \right) \pm 
\cos(\Delta m\, t) \right\} \,, \nonumber\\
I_{+-}(t) &=& g_{+}^*(t) g_{-}(t) =
-\frac{1}{2}\, e^{-\Gamma t}
\left\{ \sinh \left( \textstyle{ \frac{1}{2}} 
\Delta \Gamma\, t \right) +
i \sin(\Delta m\, t) \right\} \,, \nonumber\\
I_{-+}(t) &=& \left( I_{+-}(t) \right)^* \,.
\label{I}
\end{eqnarray}

The quantity which is directly sensitive to the interference terms is the
asymmetry 
\begin{equation}\label{5}
A(t_r,t_l) = \frac{P_\mathrm{unlike}(t_r,t_l) - P_\mathrm{like}(t_r,t_l)}
                  {P_\mathrm{unlike}(t_r,t_l) + P_\mathrm{like}(t_r,t_l)}
\end{equation}
measured in the CPLEAR experiment \cite{CPLEAR98}.
With the definition
\begin{equation}\label{def}
\rho = \frac{1}{2} \left( \left| \frac{p}{q} \right|^2 + 
                          \left| \frac{q}{p} \right|^2  \right) \,,
\end{equation}
the QM result for the asymmetry (\ref{5}) is given by
\begin{equation}\label{QMA}
A^\mathrm{QM}(t_r,t_l) = 
\frac{(1-\rho) \cosh (\frac{1}{2} \Delta \Gamma \Delta t) +
      (1+\rho) \cos (\Delta m \Delta t)}%
     {(1+\rho) \cosh (\frac{1}{2} \Delta \Gamma \Delta t) +
      (1-\rho) \cos (\Delta m \Delta t)} \,,
\end{equation}
whereas with the modified interference term the asymmetry
depends on $S$ and $\zeta$ and has to be calculated with the
full expressions (\ref{3.3}) and (\ref{3.4}), i.e.,
\begin{equation}\label{5.5}
A^S_\zeta(t_r,t_l) = 
\frac{P^S_\zeta (K^0,t_r; \bar K^0,t_l) + 
      P^S_\zeta (\bar K^0,t_r; K^0,t_l) -
      P^S_\zeta (K^0,t_r; K^0,t_l) - 
      P^S_\zeta (\bar K^0,t_r; \bar K^0,t_l)}%
     {P^S_\zeta (K^0,t_r; \bar K^0,t_l) + 
      P^S_\zeta (\bar K^0,t_r; K^0,t_l) +
      P^S_\zeta (K^0,t_r; K^0,t_l) + 
      P^S_\zeta (\bar K^0,t_r; \bar K^0,t_l)} \,.
\end{equation}

\section{Discussion of the asymmetry}

CP violation in \kkk\ mixing is small and proportional to
$\mathrm{Re}\, \varepsilon \approx (|p/q|-1)/2$. 
In the strangeness asymmetry (\ref{QMA}) calculated
according to the rules of QM, CP violation enters through the
coefficient $\rho$ (\ref{def}) and appears therefore quadratically in the
CP-violating parameter. One can easily check that the same
suppression of CP-violating effects is
true for the asymmetry (\ref{5.5}) in the bases 
$\{K^0, \bar K^0\}$ and $\{K_L, K_S\}$ corresponding to
$S=\mathbf{1}$ and $S=M$, respectively. 
Because of the smallness of CP violation in \kkk\ mixing, 
$\rho = 1$ or $(|p|^2-|q|^2)^2=0$ holds to an extremely good approximation.
Though in general the
CP-violating parameter appears linearly in Eq.(\ref{5.5}) we will
neglect CP violation in \kkk\ mixing from now on, which is
sufficient for our purpose of estimating the allowed range of
$\zeta$ for different bases. Without loss of generality we will
use the phase convention $p = q$.

As a further simplification we will restrict ourselves to
unitary matrices $S$. Since the expressions (\ref{3.3}) and
(\ref{3.4}) do not depend a phase factor multiplying $S$ we
simply assume that
\begin{equation}
S = 
\left( \begin{array}{cc} a & b \\ -b^* & a^* \end{array} \right)
\, \in \, \mathrm{SU(2)} \quad \mathrm{with} \quad
|a|^2 + |b|^2 = 1 \,.
\end{equation}
Note that with this convention the basis $\{K_L, K_S\}$ is described
by a matrix $S$ given by $a = b = i/\sqrt{2}$.

In conventional \kkk\ physics, a non-zero difference in the numbers 
of $K^0$ and $\bar K^0$ pairs
measured at $t=t_r=t_l$ signifies CP violation. Since we
assume CP conservation in \kkk\ mixing we want to ask the
question if the artificial modification (\ref{ff}) of the QM
interference terms introduces such a difference. Indeed, in
general this is the case as seen from the equation
\begin{eqnarray}
&& P^S_\zeta (K^0,t; K^0,t) - 
P^S_\zeta (\bar K^0,t; \bar K^0,t) \nonumber\\
&& = -2\, \zeta e^{-2 \Gamma t} 
\sinh \left( \frac{1}{2} \Delta \Gamma\, t \right)
\left\{ (|a|^2-|b|^2)\, \mathrm{Re}\, (ab) \cos (\Delta m\, t) +
\mathrm{Im}\, (a^2 b^2) \sin (\Delta m\, t) \right\} \,.
\label{dif}
\end{eqnarray}
However, for the bases $\{K_L, K_S\}$ with $|a|=|b|$ and $a^2$,
$b^2$ being real and $\{K^0, \bar K^0\}$ with $b=0$ the
right-hand side of Eq.(\ref{dif}) is zero (see Ref. \cite{BG98}
for an analogous consideration in the $B^0 \bar B^0$ system).

With the above simplifiying assumptions we can readily calculate
the asymmetry (\ref{5.5}) for any orthonormal basis given by a
unitary matrix $S$ and for any decoherence parameter $\zeta$.
For $\zeta = 0$ we obtain the QM result \cite{CPLEAR98}
\begin{equation}\label{QMAA}
A^\mathrm{QM}(t_r,t_l) = 
\frac{\cos(\Delta m\, \Delta t)}%
{\cosh(\frac{1}{2} \Delta \Gamma \Delta t)} \,,
\end{equation}
which is, of course, independent of $S$. For the basis $\{K_L,
K_S\}$ we arrive at the simple formula
\begin{equation}\label{6}
A^{K_LK_S}_\zeta (t_r,t_l)= (1-\zeta) A^\mathrm{QM}(t_r,t_l) \,,
\end{equation}
whereas for $\{K^0, \bar K^0\}$ we obtain
\begin{eqnarray}\label{7}
A^{K^0 \bar K^0}_\zeta (t_r,t_l)
&=&\frac{\cos(\Delta m \Delta t) - \frac{1}{2} \zeta
\{ \cos(\Delta m \Delta t)-\cos(\Delta m (t_r+t_l)) \} }%
{\cosh(\frac{1}{2} \Delta \Gamma \Delta t) - \frac{1}{2} \zeta 
\{ \cosh(\frac{1}{2} \Delta \Gamma \Delta t) -
\cosh(\frac{1}{2} \Delta \Gamma (t_r+t_l)) \} } \,.
\end{eqnarray}
Eqs.(\ref{6}) and (\ref{7}) represent the two cases for which we
will perform a numerical estimate of the decoherence parameter
$\zeta$ in the next section.

In our formalism, Furry's hypothesis (``method A'' in Ref.
\cite{furry}) corresponds to $\zeta = 1$. Looking at Eqs.(\ref{6}) and
(\ref{7}) we read off in this case
\begin{eqnarray}
A^{K_LK_S}_1 (t_r,t_l) & = & 0 \,, \label{klks} \\
A^{K^0 \bar K^0}_1  (t_r,t_l) & = &
\frac{\cos(\Delta m \Delta t) + \cos(\Delta m (t_r+t_l))}%
{\cosh(\frac{1}{2} \Delta \Gamma \Delta t) +
\cosh(\frac{1}{2} \Delta \Gamma (t_r+t_l)) }
\,. \label{k0k0}
\end{eqnarray}
In the basis $\{K_L, K_S\}$ the asymmetry vanishes \cite{CPLEAR98}, 
whereas in the basis $\{K^0, \bar K^0\}$
spontaneous factorization leads to a non-zero asymmetry (\ref{k0k0}).
One might ask the question for
which bases the asymmetry (\ref{5.5}) is zero at $\zeta = 1$ for
arbitrary $t_r$, $t_l$. The answer to this question is given by the
following proposition.\\[2mm]
\textbf{Proposition:} \textit{With the assumptions $S \in$ SU(2) and
$|p/q|=1$, Furry's hypothesis leads to a zero asymmetry (\ref{5.5}) if and
only if spontaneous factorization of the \kkk\ wave function
takes place in the $\{K_L, K_S\}$ basis.}\\[2mm]
\emph{Proof:} We study the general expression
\begin{eqnarray}
&&
P^S_\zeta (K^0,t_r; \bar K^0,t_l) + P^S_\zeta (\bar K^0,t_r; K^0,t_l) -
P^S_\zeta (K^0,t_r; K^0,t_l) - P^S_\zeta (\bar K^0,t_r; \bar K^0,t_l)
\nonumber\\
&& \hphantom{AAAAA}
 = e^{-\Gamma (t_r+t_l)}  \left\{ \vphantom{\frac{\alpha}{\alpha}}
\cos (\Delta m \Delta t) \right. 
\nonumber\\
&& \hphantom{AAAAABBBBaaaaa} -\frac{1}{2} \zeta \left[ 
\left( |(a^*)^2 + b^2|^2 + 4|a|^2|b|^2 \right) \cos (\Delta m \Delta t) 
\right. \nonumber\\
&&
\hphantom{AAAAABBBBaaai -\frac{1}{2}\, \zeta }
-\left( |(a^*)^2 + b^2|^2 - 4|a|^2|b|^2 \right) \cos (\Delta m (t_r+t_l)) 
\nonumber\\
&& 
\hphantom{AAAAABBBBaaai -\frac{1}{2} \zeta } 
\left. \left.
-4\, (|a|^2-|b|^2)\, \mathrm{Im}\, (ab) \sin (\Delta m (t_r+t_l))
\right] \vphantom{\frac{\alpha}{\alpha}} \right\} \,.
\end{eqnarray}
The right-hand side of the equation is zero $\forall\: t_r, t_l$
at $\zeta=1$ if the system of equations
\begin{eqnarray}
|(a^*)^2 + b^2|^2 + 4|a|^2|b|^2 & = & 2 \,, \nonumber\\
|(a^*)^2 + b^2|^2 - 4|a|^2|b|^2 & = & 0 \,, \nonumber\\
(|a|^2-|b|^2)\, \mathrm{Im}\, (ab) & = & 0
\end{eqnarray}
is fulfilled. The general solution of this system is given by
$a=e^{i\alpha}/\sqrt{2}=\epsilon b^*$ with 
$\epsilon = \pm 1$ and $\alpha$ being an arbitrary phase. It
is then easy to show that a matrix $S$ with these coefficients
represents the basis $\{K_L, K_S\}$ apart from trivial
redefinitions. $\Box$ 

\section{Estimation of the decoherence parameter $\zeta$ 
from the data of the CPLEAR Experiment}

In the CPLEAR experiment \cite{CPLEAR98} at CERN, \kkk\ pairs are 
produced by $p \bar p$ 
annihilation at rest. These pairs are predominantly in an antisymmetric state
with quantum numbers 
$J^{PC}=1^{--}$ and the strangeness of the kaons is detected 
via strong interactions in surrounding absorbers.
The experimental set-up has two configurations. In
configuration C(0) both kaons have nearly equal proper times 
($t_r \approx t_l$) when they interact in the absorber. 
This fulfills the conditions for an EPR-type experiment. In
configuration C(5) the flight-path difference is 5 cm on average,
corresponding to a proper time difference 
$| t_r-t_l | \, \approx 1.2 \,\tau_{S}$.

The asymmetry (\ref{5}) is measured for these two configurations, 
giving the experimental results \cite{CPLEAR98}
$A^\mathrm{exp}(0) = 0.81\pm 0.17$ for C(0) and
$A^\mathrm{exp}(5) = 0.48\pm 0.12$ for C(5).
On the other hand, the QM predictions for the asymmetry, 
when corrected according to the
specific experimental design, yield
$A^\mathrm{QM}_\mathrm{corr}(0) = 0.93$ and
$A^\mathrm{QM}_\mathrm{corr}(5) = 0.56.$
These values are in agreement with the above experimental ones,
demonstrating in this way the non-separability of the QM \kkk\ wavefunction.

Our approach is somewhat different.
We estimate the amount of decoherence and show,
as already emphasized, its dependence on the basis chosen. Specifically, we fit
the decoherence parameter $\zeta$ by comparing the asymmetry 
for the two cases (\ref{6}) and (\ref{7}) 
with the experimental data of CPLEAR. For the fit of the parameter 
$\zeta$ we use
the least squares method of Ref. \cite{orear} which is an effective 
variance method,
taking into account not only the experimental error in 
$A(t_r,t_l)$ but also the variations
in $x_{r,l} = 4.28\: \mathrm{cm}\: (t_{r,l}/\tau_{S})$, 
the space coordinates where the strangeness of the kaons is measured.

First we discuss the asymmetry in the $\{K_L,K_S\}$ basis where
formula (\ref{6}) is relevant. This case is rather easy to handle
because $\zeta$ enters into the asymmetry linearly. Our 
fit result is
\begin{equation}\label{fit1}
\bar \zeta = 0.13 
\begin{array}{cc}
+0.16\\
-0.15
\end{array} \,.
\end{equation}
The CL for this fit is 97\%.

In Fig.\ 1 we have plotted the asymmetry (\ref{6}) for the values 
of $\zeta$ given in Eq.(\ref{fit1}). The three solid lines 
represent the asymmetry (\ref{6}) for the meanvalue $\bar \zeta$
and its one standard deviation values $\bar\zeta = 0.29$ and
$-0.03$ (see Eq.(\ref{fit1})), respectively.
The dashed curve shows the QM prediction (\ref{QMAA}) ($\zeta = 0$). 
The experimental results for the two configurations C(0) and C(5) are
also indicated according to Fig.\ 9 in Ref. \cite{CPLEAR98}, where 
in comparison to the experimental numbers quoted above some
``background subtraction'' has been performed \cite{CPLEAR98}.
We have scaled the variable $\Delta t$ in the QM asymmetry
(\ref{QMAA}) in order to reproduce the QM curve
in Fig.\ 9 of the CPLEAR Collaboration \cite{CPLEAR98}. 

The case of the asymmetry in the $\{K^0, \bar K^0 \}$
basis (\ref{7}) is a bit more intricate than the previous one 
because formula (\ref{7}) depends not only
on the time difference $\Delta t = t_r - t_l$ but also on 
the sum $t_r + t_l$, and, moreover, the dependence on $\zeta $ is not 
linear. We estimate the values of the sum $t_r + t_l$
and their variations from the experimental
configurations C(0) and C(5). 
With this additional but less accurate information
we obtain the estimate
\begin{equation}\label{fit2}
\bar \zeta \sim \, 0.4 \pm 0.7 .
\end{equation}
Here the quality of the fit is also good with a CL of about 67\%.

In order to assess the validity of QM there are two measures 
associated with the parameter
$\zeta$. They are given by the distance of the fitted mean value from
0 and from 1, each expressed in units of one standard deviation, 
corresponding to pure QM and to Furry's hypothesis, respectively.
Considering first the distance of $\bar \zeta$ from 0, we see that for
both bases, $\{K_L,K_S\}$ and $\{K^0, \bar K^0 \}$, the results of our
fit are very well compatible with QM.
On the other hand, $\zeta = 1$ corresponds to Furry's hypothesis, i.e.,
total decoherence or spontaneous factorization.
In this case,
with the $\{K_L,K_S\}$ basis the asymmetry vanishes identically (see
Eq.(\ref{klks})), which -- as found by
the CPLEAR Collaboration \cite{CPLEAR98} -- has a probability of 
less than $10^{-4}$. In the
$\{K^0,\bar K^0 \}$ basis, however, formula (\ref{k0k0}) gives the 
values 
\begin{equation}
A^{K^0 \bar K^0}_1(0) = 0.90, \quad
A^{K^0 \bar K^0}_1(5) = 0.50  \nonumber
\end{equation}
for the two configurations C(0) and C(5), respectively,
which are within one standard deviation of the corresponding data 
and therefore not excluded at all.
Note also that the two fit results for $\zeta$ (\ref{fit1}) and
(\ref{fit2}) clearly disfavour negative values of the decoherence
parameter. 

Last but not least we want to mention that we also 
checked the effect of the experimental
errors of the input quantities $\Delta m$ and $\Delta \Gamma$ 
on the asymmetries (\ref{6}) and (\ref{7}).
The resulting error in the asymmetries is less than 1.5\% 
for all times. Therefore, we can safely neglect these errors
for our purpose.

\section{Conclusions}

We reconsidered the results of the experiment of the CPLEAR Collaboration 
\cite{CPLEAR98} which investigated
strangeness correlations of \kkk\ pairs produced by $p\bar p$ 
annihilation. We introduced the decoherence parameter $\zeta$ in order 
to measure quantitatively deviations from pure
QM. This introduces a certain arbitrariness since the quantity
we consider, the asymmetry
(\ref{5.5}), depends strongly on the basis chosen (pure QM, of course,
not). This is not so surprising,
it is merely the consequence of the freedom in QM to choose a basis in 
$K^0$--$\bar K^0$ space,
which eventually leads to different interference terms for different 
basis choices \cite{dass,BG98}. Thus the
parameters $\zeta$ modifying these different interference terms will 
get in general different
values when confronted with experiment. What is essential, however, 
is the existence of a
basis where the \kkk\ system is far away from total decoherence and 
the corresponding $\zeta$
is close to 0 in agreement with QM. We have shown that the 
``best basis'' in this respect is the $\{K_L,K_S\}$ basis (see proposition).

Furthermore, because of the introduction of the matrix S, 
which is arbitrary apart from
$S\in$ SU(2), even for $|p/q| = 1$ we have in general 
$P^S_\zeta (K^0,t; K^0,t) - P^S_\zeta (\bar K^0,t; \bar K^0,t) \ne 0$,
a sign for CP violation in \kkk\ mixing. 
Thus, in our ad hoc modification of QM by 
the decoherence parameter $\zeta$, the matrix $S$ mimics CP violation
in general. However, as we have shown, for the
two bases under consideration, $\{K_L, K_S\}$ and $\{K^0, \bar K^0\}$,
the difference (\ref{dif}) vanishes. This is analogous to considerations 
in the $B^0 \bar B^0$ system \cite{BG98}.

Finally, we want to mention that our statistical analysis is not
very refined and, in addition, further experimental 
data could diminish the errors of our fits.
Nevertheless, long range interference effects, i.e., the presence of the 
QM interference term in the \kkk\
system, are quite well confirmed in agreement with QM 
and thus also interference effects
of massive particles over macroscopic distances.

\begin{figure}
\scalebox{2.8}{\includegraphics*[-0.3cm,-0.5cm][12cm,3.5cm]{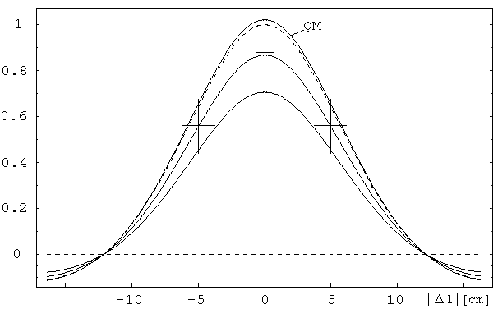}}
\caption{The asymmetry (\ref{6})
as a function of the difference in
the distances travelled by the kaons to the points where their
strangeness is measured. The dashed curve corresponds to QM with the
decoherence parameter $\zeta = 0$, whereas the solid curves correspond
to the values of $\zeta$ obtained by the fit (\ref{fit1}) to the
CPLEAR data. The two data points represented by
the crosses have been taken from Fig.\ 9 of Ref. \protect\cite{CPLEAR98}. 
The horizontal dashed line indicates the zero asymmetry for $\zeta = 1$ 
(\ref{klks}), the consequence of Furry's hypothesis with respect to
the $\{ K_L,  K_S \}$ basis.}
\end{figure}

\end{document}